\documentclass[1p,twocolumn]{elsarticle} 
\usepackage[fleqn]{amsmath}
\usepackage{amssymb}
\usepackage{bm}
\usepackage{graphicx}
\usepackage{indentfirst}
\usepackage{subfigure}
\usepackage{verbatim}
\usepackage{caption}
\usepackage{color}
\usepackage{float}
\biboptions{sort&compress}

\usepackage[normalem]{ulem}  
\usepackage{color} 
\usepackage{xcolor}

\renewcommand{\sout}{\bgroup \color{red} \ULdepth=-.5ex \ULset}
\newcommand{\soug}{\bgroup \color{cyan} \ULdepth=-.5ex \ULset}
\begin{document}
\title{Heavy quarkonium dissociation in the finite space of heavy-ion collisions}
\author{Jihong Guo}
\author{Wu-Sheng Dai}
\author{Mi Xie}
\author{Yunpeng Liu}
\ead{yunpeng.liu@tju.edu.cn}
\address{Department of Physics, Tianjin University, Tianjin 300350, P. R. China}
\begin{abstract} 
   The dissociation of heavy quarkonia in the constrained space is calculated at leading order compared with that in infinitely large medium. To deal with the summation of the discrete spectrum, a modified Euler-Maclaurin formula is developed as our numerical algorithm. We find that with the constraint in space, the dissociation  of quarkonia at early time becomes negligible.
\end{abstract}
\maketitle
\section{Introduction}
The quark-gluon plasma (QGP) is believed to be the state of quark matter at extremely high temperature and/or extremely high density. Such conditions can be found in laboratory only by relativistic heavy-ion collisions. The volume in which the QGP is produced is at the same scale as that of a nucleus, and the QGP can-not be detected directly. Heavy quarkonia are important probes of the QGP produced in heavy-ion collisions, since they suffer suppression in the QGP and almost survive the hadron gas. Fruitful results are obtained in experiments
~\cite{Abreu:1999qw, Adare:2006ns, Adamczyk:2012ey, Chatrchyan:2012np, Abelev:2012rv, Adare:2011yf} including the nuclear modification factors of quarkonia at different energies, rapidities, and transverse momenta. On the other hand, different models are put up to calculate the suppression. The most early idea is that different excited states of quarkonia melt in the QGP sequentially due to the color screening~\cite{Matsui:1986dk, Blaizot:1996nq}. A different point of view~\cite{BraunMunzinger:2000px} attributes all the observed heavy quarkonia to the thermal balance between the open and hidden heavy flavors. Meanwhile, the calculation based on scattering cross section was put up~\cite{Xu1996, Oh:2000qr, Grandchamp:2001pf, Zhuang:2003fu}, and the regeneration of quarkonia from heavy quarks in the QGP is considered~\cite{Thews:2000rj, Grandchamp:2001pf, Yan:2006ve}. The properties of heavy quarkonia have also been studied in the framework of effective field theory~\cite{Brambilla:1999xf} and lattice QCD~\cite{Asakawa:2003re}. The theories are still under development recent years~\cite{Zhao:2007hh, Riek:2010fk, Uphoff:2010sh, Grigoryan:2010pj, Song:2011ev, Liu:2012zw, Liu:2013kkg, Song:2013lov, Zhou:2014kka, Ding:2015ona, Chen:2016vha, Shi:2017qep, Chen:2018dqg}.
\par
All above theories focus on the properties of quarkonia in infinitely large medium, while the volume of the QGP  is finite in experiments, especially at early time after the collision. One direct consequence is that the  spectrum of gluons becomes different in the finite space  compared with that in the infinite space at the same temperature, and therefore the dissociation rate of heavy quarkonia differs. In this letter we will discuss the corresponding effect. Note that the longitudinal size of the medium is much smaller than that in the transverse directions, we will assume that the medium is infinitely large in the transverse directions. For simplicity, we only consider the initially produced quarkonia at middle rapidity and take leading order cross section of the gluons dissociation process.
\par
In Section \ref{se_tm}, we introduce the model to describe the  suppression of quarkonia, where a summation in the spectrum of gluons in the finite space is introduced to replace the integral in the infinite space. To deal with the summation, a modified Euler-Maclaurin Formula is developed in Section \ref{se_mEM} as our numerical algorithm. Results of the gluon spectrum and the dissociation rate of quarkonia in the finite space compared with that in the infinite space is shown in Section \ref{se_nr}. The effective initial time is also discussed in this section. A short conclusion is given in Section \ref{se_cl}. We take the natural units $\hbar=c=k_{\rm B}=1$. A pair of square brackets [~] in an equation within this letter is always used as a floor function.
\section{Dissociation of quarkonia}    \label{se_tm}
In high energy nuclear collisions, the distribution function $f_{H}(\bm{p},\bm{x},t)$ of heavy quarkonia $H$\ at $(\bm{p},\bm{x})$ in the phase space at time $t$ is controlled by the equation~\cite{zhu:2004nw}
\begin{eqnarray}   
  \partial f_{H}/\partial t&=&-\alpha_{H}f_{H},
\end{eqnarray} 
at middle rapidity in the lab frame, where $\alpha_{H}(\bm{p},\bm{x},t)$ is the dissociation rate of $H$ in the hot medium. We have neglected both the leakage effect~\cite{zhu:2004nw} and the mean field effect~\cite{chen:2012gg} on heavy quarkonia.
\par Before discussing the loss term $\alpha_{H}$ in the finite space, we first write it in the infinite space. For simplicity, we only consider the gluon dissociation process $g+H\to Q+\bar{Q}$\ in the QGP phase, and the loss term $\alpha_{H}$ is
\begin{eqnarray}     
  \begin{aligned}   
  \label{lose1}
  \alpha_{H}(\bm{p},\bm{x},t) &=\frac{1}{2E_{H}}\int\frac{{\rm d}^{3}\bm{k}}{(2\pi)^{3}2E_{g}}W^{Q\bar{Q}}_{gH}(s)f_{g}(\bm{k},\bm{x},t),
\end{aligned}
\end{eqnarray} 
where $E_{H}$\ and $E_g$\ are the energies of the heavy quarkonium $H$ and the gluon, respectively, in the lab frame. The transition probability $W_{gH}^{Q\bar{Q}}(s)=4\sigma_{gH}\sqrt{(p^{\mu}k_{\mu})^{2}-m^{2}_{H}m^{2}_{g}}$ is a function of $s=(p_{\mu}+k_{\mu})^{2}$ with $p_{\mu}$ and $k_{\mu}$ being the four-momenta of $H$ and the gluon, respectively. The gluon mass $m_{g}$ is taken zero. The dissociation cross section~\cite{Peskin:1979va,Bhanot:1979vb} is
 \begin{eqnarray}    
  \sigma_{gH}(\omega)&=&A_{0}\frac{(\omega/\epsilon_{H}-1)^{3/2}}{(\omega/\epsilon_{H})^{5}},
\end{eqnarray} 
with $A_{0}=2^{11}\pi/(27\sqrt{m_{Q}^{3}\epsilon_{H}})$, where $\omega = \left(s-m_{H}^{2}\right)/2m_{H}$ is the gluon energy in the rest frame of $H$. The binding energy  is replaced by the threshold energy $\epsilon_H=(4m_{Q}^{2}-m_{H}^{2})/(2m_{H})$\ in our calculation in order to include the recoiling effect ~\cite{liu:2009nb}, where $m_{Q}$ is the mass of a heavy quark. The distribution function of gluons is assumed to be thermal
\begin{eqnarray}     
  \label{gluon distribution}
  f_{g}(\bm{k},\bm{x},t)&=&\frac{N_{g}}{e^{u^{\mu}k_{\mu}/T}-1},
\end{eqnarray}
where $T(\bm{x},t)$ and $u^{\mu}(\bm{x},t)$ are the local temperature and four-velocity, respectively, and $N_{g}=16$\ is the degeneracy of gluons. The dissociation in the hadron phase is neglected.
\par For simplicity, we describe the fireball by Bjorken's hydrodynamics, which neglects the transverse flow of the medium, and the entropy density is inversely proportional to time. For the spatial distribution, the entropy is assumed to be proportional to the number density of participants $n_{\rm{part}}$. We take the equation of state as that of ideal parton gas. Then we have
\begin{eqnarray}
  \label{T0} 
  T({\bm x}, t)&=& T^*\left(\frac{t^*}{t}\frac{n_{\rm{part}}(\bm{x})}{n_{\rm{part}}(\bm{x}={\bm 0})}\right)^{\frac{1}{3}},
  \label{np}
\end{eqnarray}    
where $T^*$\ is the temperature at ${\bm x}={\bm 0}$ and $t=t^*$.
The number density  of binary collision $n_{\rm part}$\ is calculated by the Glauber model~\cite{Miller:2007ri} with the Woods-Saxon density profile
  $\rho(\bm{r})=\frac{\rho_{0}}{1+e^{(r-r_{0})/a}}$.
The specific parameters of  ${ }^{197}{\rm Au}$ ($r_{0}=6.38$ fm, $a=0.535$ fm and $\rho_{0}=0.169\ \rm{fm}^{-3}$) used in the numerical calculations are from Ref.~\cite{deJager:1974liz}.
\par Now we consider the loss term $\alpha_{H}$ in the finite space. In relativistic heavy-ion collisions, the QGP only exists in a small region in space, especially at early time after the collision when the longitudinal size is small. For simplicity, we assume that the fireball is infinitely large in the transverse directions and the longitudinal size of the fireball is $L=2t$ at time $t$ after the collision.
The eigen energy of a gluon in the lab frame is $E_g =\sqrt{k_{T}^{2}+\left(\frac{n\pi}{L}\right)^{2}}$, where $k_{T}$ is the transverse momentum of the gluon and $n=1,2\cdots$ is the quantum number of $k_{z}$. The loss term in Eqn.~(\ref{lose1}) is replaced by
\begin{eqnarray}  
  \begin{aligned}
  \label{lose2}
  \alpha_{H}&=\frac{1}{2E_{H}L}\sum_{n}\int\frac{{\rm d}^{2}\bm{k}_{T}}{(2\pi)^{2}2E_{g}}W_{gH}^{Q\bar{Q}}(s)
  f_{g}(\bm{k}_{T},n,\bm{x},t),
\end{aligned} 
\end{eqnarray}  
where $f_{g}(\bm{k}_{T},n,\bm{x},t)$ takes the same form as in Eqn.~(\ref{gluon distribution}) with $k_{z}=n\pi/L$. Note that Eqn.~(\ref{gluon distribution}) is invariant under a transverse boost.  Eqn.~(\ref{lose2}) can be rewritten in the quarkonium frame as
\begin{eqnarray}
  \label{loss}
  \alpha_{H}=\frac{1}{4E_{H}}\int \frac{{\rm d}\omega}{\omega} f_{\omega}W_{gH}^{Q\bar{Q}}(s).
\end{eqnarray}  
The number density $f_{\omega}$ of gluons in unit energy is 
\begin{eqnarray} 
  \label{fomega}
  f_{\omega}&=&\frac{ {\rm d}N}{ {\rm d}\omega{\rm d}V}
  =\frac{1}{L}\sum_{n}\int\frac{ {\rm d}^{2}{\bm k}_{T}^{'}}{(2\pi)^{2}{\rm d}\omega}f_{g}({\bm k}_{T}^{'},n,{\bm x},t),
\end{eqnarray} 
with the gluon thermal distribution function
\begin{eqnarray}
  f_{g}(\bm{k}_{T}^{'},n,\bm{x},t)=\frac{N_{g}}{e^{u^{\mu}_{H}k^{'}_{\mu}/T}-1},
\end{eqnarray}
where $u^{\mu}_{H}$ and $k^{'}_{\mu}$ are, respectively,  the four-velocity and four-momentum in the quarkonium frame.
\section{Modified Euler-Maclaurin Formula} \label{se_mEM} 
In order to work out the summation in Eqn.~(\ref{fomega}), we develop a modified Euler-Maclaurin formula. The original Euler-Maclaurin formula~\cite{TomM} is
\begin{eqnarray} 
  \label{Euler}
  \sum_{i=a}^{b}f(i)=\int_{a}^{b}f(x){\rm d}x+\frac{f(b)+f(a)}{2}
  +\sum_{r=1}^{[\frac{n}{2}]}\frac{B_{2r}}{(2r)!}\left(f^{(2r-1)}(b)-f^{(2r-1)}(a)\right)
  +R_{n},
\end{eqnarray}   
where $B_{2r}$ is the (2r)th Bernoulli number~\cite{Abramowitz}. The remainder term is
\begin{eqnarray}   
  R_{n}=\frac{(-1)^{n+1}}{n!}\int_{a}^{b}f^{(n)}(x)P_{n}(x){\rm d}x,
\end{eqnarray} 
where $P_{n}(x)$ is the periodic Bernoulli polynomial~\cite{Lehmer}.  
\par 
Sometimes the first a few terms are important (e.g. low energy states in calculating the partition function of bosons at low temperature). Thus we take the summation of the first $m$ terms explicitly. The Bernoulli number $B_{2r}$ grows fast with $r$, and the remainder term $R_n$\ often diverges as $n\rightarrow\infty$. The Fourier series of $P_{n}$ is~\cite{Luo2010Fourier} 
\begin{eqnarray}
  \label{pn}
P_{n}(x)=-n!\sum_{k\in Z-\{0\}}\frac{e^{2\pi ikx}}{(2\pi ik)^{n}}.
\end{eqnarray}  
We take $2p$ terms with $|k|\leqslant p$ in Eqn.~(\ref{pn}) and leave the others to the new reminder term $R_{mnp}$. Then the modified Euler-Maclaurin formula is 
\begin{eqnarray} 
  \label{m-n}
    \sum_{i=a}^{b}f(i)&=&\sum_{i=a}^{a'-1}f(i)+\frac{f(b)+f(a')}{2}
    +\int_{a'}^{b}f(x)\frac{ {\rm sin}(2p+1)\pi x}{ {\rm sin}\pi x}{\rm d}x\nonumber\\
    &&+\sum_{r=1}^{[\frac{n}{2}]}\frac{2(-1)^{r-1}}{(2\pi)^{2r}}S_{2r,p}M_{2r-1}(b,a')
    +R_{mnp},
\end{eqnarray}      
with $S_{2r,p}=\zeta(2r)-\sum\limits_{k=1}^{p}\frac{1}{k^{2r}}$, $M_{2r-1}(b,a')=f^{(2r-1)}(b)-f^{(2r-1)}(a')$, and $a'=a+m$. Here the $\zeta$\ in $S_{2r,p}$\ is the Riemann zeta function. Dropping the new remainder term 
\begin{eqnarray} 
    R_{mnp}=2\sum_{k=p+1}^{\infty}\int_{a'}^{b} f(x){\rm cos}(2\pi kx){\rm d}x
    +\sum_{r=1}^{[\frac{n}{2}]}\frac{2(-1)^{r}}{(2\pi)^{2r}}S_{2r,p}M_{2r-1}(b,a'),
\end{eqnarray}    
a m-n-p cut of the modified Euler-Maclaurin formula is obtained, which can be used as a numerical algorithm of the original summation. Any accuracy can be achieved by choosing $n$\ and $p$.
In practice, the integral in Eqn.~(\ref{m-n}) can also be calculated by  
\begin{eqnarray} 
     \int_{a^{'}}^{b}f(x)\frac{ {\rm sin}(2p+1)\pi x}{ {\rm sin}\pi x}{\rm d}x
     = 2\sum_{k=1}^{p}\int_{a^{'}}^{b} f(x){\rm cos}(2\pi kx){\rm d}x+\int_{a^{'}}^{b} f(x){\rm d}x.
\end{eqnarray}  
\par 
The loss term in Eqn.~(\ref{loss})  is calculated by a 2-2-1 cut of the  modified Euler-Maclaurin formula as
\begin{equation}
  \label{sim}
   \alpha_{H}=\sum_{i=1}^{2}H(i)+\int_{3}^{\infty}H(y)\frac{ {\rm sin}3\pi y }{ {\rm sin}\pi y}{\rm d}y+\frac{1}{2}H(3)-\frac{2}{(2\pi)^{2}}(\zeta(2)-1)H^{(1)}(3)
\end{equation}   
in the next section with
\begin{eqnarray}
  \begin{aligned}
     H(y)&=\frac{1}{4E_{H}L}\int \frac{{\rm d}^{2}\bm{k}_{T}^{'}}{(2\pi)^2\omega}W_{gH}^{Q\bar{Q}}(s)f_{g}(\bm{k}_{T}^{'},y,\bm{x},t).
\end{aligned}
\end{eqnarray} 
\section{Numerical Results}\label{se_nr}
  Now we discuss the number density $f_{\omega}$ in unit energy in Eqn.~(\ref{fomega}) of gluons which is called density in the following for short. 
\begin{figure}[!hbt]    
  \centering
    \includegraphics[width=3.1in,height=2.45in]{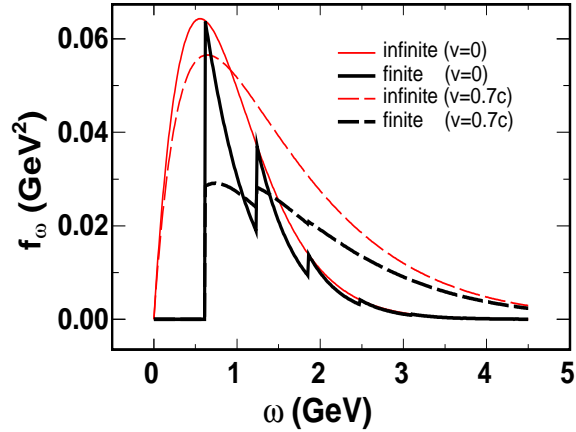}
  \caption{(Color online) The gluon number density $f_{\omega}$\ in unit energy in the finite space ($L=1$ fm) and the infinite space ($L=\infty$) at $T=0.35$ GeV with $v=0$\ and $v=0.7c$.}
  \label{fig.1}
\end{figure}
Fig.~\ref{fig.1} shows the density as a function of $\omega$ in a static ($v=0$) or moving frame ($v=0.7c$) in a finite fireball ($L=1$ fm) compared with that in an infinite fireball ($L=\infty$) at $T=0.35$ GeV. The density in a finite fireball is never larger than that in an infinite fireball， and the gluons whose energy is less than the ground state energy $\omega_0=\pi/L$ vanish. In order to understand the properties of $f_{\omega}$, we consider two limits: in the static frame and in the fast moving one. (i) In the static frame ($v=0$), Eqn.~(\ref{fomega}) can be simplified as  
\begin{eqnarray}
  f_{\omega}^{L}=\frac{N_{g}}{2\pi L}\left[\frac{\omega}{\omega_{0}}\right]\frac{\omega}{e^{\omega/T}-1},
\end{eqnarray}  
while the density in the infinite space is
\begin{eqnarray}
  f_{\omega}^{\infty}=\frac{N_{g}}{2\pi^{2}}\frac{\omega^2}{e^{\omega/T}-1}.
\end{eqnarray}  
The ratio $f_r=f_{\omega}^{L}/f_{\omega}^{\infty}$ satisfies $(1-\frac{\omega_{0}}{\omega})\leqslant f_{r}\leqslant 1$ at $\omega\geqslant\omega_{0}$ and the equality holds only at $\omega=n\omega_{0}$\ $(n=1,2,3\cdots)$\ as shown in Fig.~\ref{fig.1}. (ii) In the fast moving frame, Eqn.~(\ref{fomega}) can be simplified in the condition of both $|u^{T}_{H}|\gg 2T\omega/\omega_{0}^{2}$ and $|u^{T}_{H}|\gg T/\omega$ as
\begin{eqnarray}  
  \label{fomegaf}
  f_{\omega}^L=\frac{N_g}{L}\sqrt{\frac{\omega}{(2\pi)^{3}B}}e^{-A\omega+B\omega\sqrt{1-(\frac{\omega_{0}}{\omega})^{2}}},
\end{eqnarray}
with $A=u^{0}_{H}/T$ and $B=|u^{T}_{H}|/T$. In the infinite space, the density is 
  \begin{eqnarray}
    \label{finf}
    f_{\omega}^{\infty}=\frac{N_g\omega}{(2\pi)^{2}B}e^{(-A+B)\omega}.
  \end{eqnarray}
  The ratio in the fast moving frame is
  \begin{eqnarray} 
    \label{fr}
    f_{r} &=&\frac{1}{L}\sqrt{\frac{2\pi B}{\omega}}e^{B\omega\left(\sqrt{1-(\frac{\omega_{0}}{\omega})^{2}}-1\right)}\ll1,
\end{eqnarray}
which shows strong suppression of the density in the finite space.
\par The change of the density leads to the change of the loss term $\alpha_{H}$ defined in Eqn.~(\ref{loss}). 
\begin{figure}[!htb]   
  \centering
      \includegraphics[width=3.1in,height=2.45in]{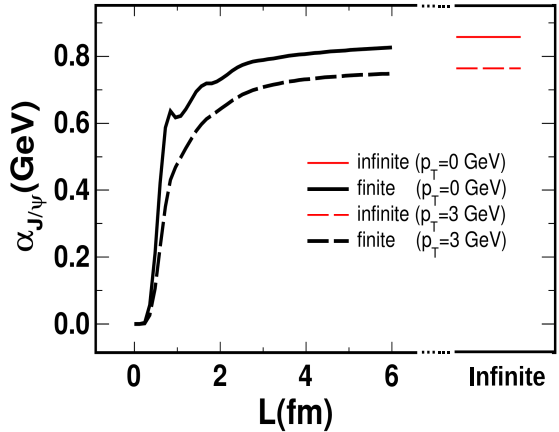}
\caption{(Color online) The loss term $\alpha_{J/\psi}$ as a function of longitudinal size $L$ at transverse momentum $p_{T}=0$ and $3$ GeV at $T=0.35$\ GeV.}
  \label{fig.2}
\end{figure} 
As a result of the discussion in the previous paragraph, in the static frame, the loss term $\alpha^{L}$  in the finite space lies between $\frac{1}{2}\alpha^{\infty}$ and $\alpha^{\infty}$ at $\omega_{0}<\epsilon_H$, and it is far smaller than $\alpha^{\infty}$ at $\omega_{0}\gg\epsilon_H$ with $\omega_{0}\gtrsim T$. In the following we discuss the dissociation rate of $J/\psi$ in two cases. (i) We fix the temperature $T=0.35$ GeV as a constant. Fig.~\ref{fig.2} shows the loss term $\alpha_{J/\psi}$ as a function of $L$ at the transverse momentum $p_{T}=0$ and $3$ GeV of $J/\psi$. The finite system approaches to the infinite system when $L$ is large, and the finite volume effect is remarkable at a small $L$($\lesssim\frac{\pi}{\epsilon_{J/\psi}}$). The kinks of the line in Fig.~\ref{fig.2} come from the jumps of $f_{\omega}$ in Fig.~\ref{fig.1}.  (ii) We evolve temperature $T$ according to the Bjorken's hydrodynamics. In Fig.~\ref{fig.3},
\begin{figure}[!hbt]
  \centering
    \includegraphics[width=3.1in,height=2.45in]{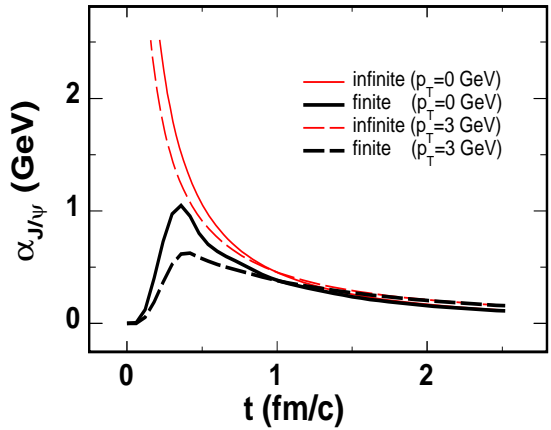}
\caption{(Color online) The loss term $\alpha_{J/\psi}$ as a function of the time $t$ in the finite space and the infinite space with the transverse momentum $p_{T}=0$ and $3$ GeV of $J/\psi$ with Bjorken's hydrodynamics.}
\label{fig.3}
\end{figure}      
we show the $\alpha_{J/\psi}$ as a function of time $t$ with transverse momentum $p_{T}=0$ and $3$ GeV with $T^*=0.35$ GeV at $t^*=0.6$ fm/$c$. In the infinite space, the loss term is divergent at $t= 0$. This divergence is usually avoided by constraining the suppression after the formation of the QGP and $J/\psi$. However our calculation indicates that, even if the formation times of the QGP and $J/\psi$ are early enough, the loss term $\alpha_{J/\psi}$ is still negligible at small $t$. The results of $\Upsilon(1s)$ are similar.
\par
\begin{figure}[!hbt]
  \centering
  \includegraphics[width=3.1in,height=2.45in]{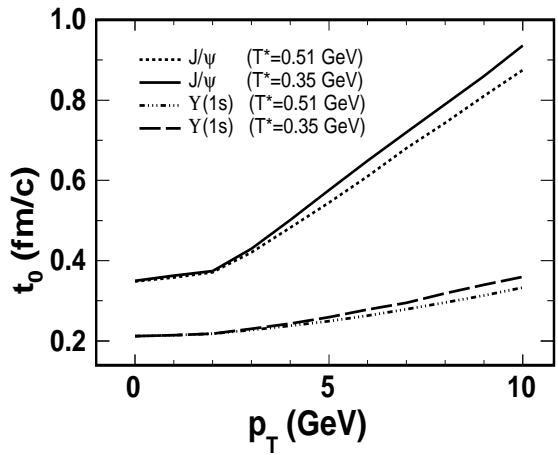}
  \caption{The effective initial time $t_{0}$ of $J/\psi$\ and $\Upsilon(1s)$\ as a function of $p_{T}$ with Bjorken's hydrodynamics with $T^*=0.35$\ and $0.51$\ GeV, respectively,  at $t^*=0.6$ fm/$c$.}
    \label{fig.4}
 \end{figure}

 We define the effective initial time $t_{0}$ by requiring the suppression in the finite space since $t=0$ is equal to that in the infinite space since $t=t_{0}$ which should be an additional cut besides the ordinary two formation times as mentioned above. The calculated effective initial time $t_{0}$ of $H (J/\psi,\Upsilon(1s))$ as a function of $p_{T}$ with temperature $T^*=0.35$ and $0.51$ GeV at $t^{*}=0.6$ fm/$c$ is shown in Fig.~\ref{fig.4}.
 The threshold energy $\epsilon_{H}$ is important to the effective initial time. As discussed in previous paragraph, in the static frame, we have $\alpha^{L}\sim\alpha^{\infty}$ at $\omega_{0}\ll\epsilon_H$, and $\alpha^{L}\ll\alpha^{\infty}$ at $\omega_{0} \gg\epsilon_H$ with $\omega_{0}\gtrsim T$. Therefore, $\omega_{0}=\epsilon_H$ gives a rough estimate of $t_{0}$, which leads to the relation $t_{0_{J/\psi}}/t_{0_{\Upsilon(1s)}}\approx\epsilon_{\Upsilon(1s)}/\epsilon_{J/\psi}$. In our calculation, the ratio of $\epsilon_{\Upsilon(1s)}/\epsilon_{J/\psi}$ and $t_{0_{J/\psi}}/t_{0_{\Upsilon(1s)}}$ are $1.67$ and $1.65$, respectively. At high-$p_{T}$, the suppression of $f_{\omega}$\ is strong according to Eqn.~(\ref{fr}). Therefore $t_0$ increases monotonically with $p_T$. No strong dependence of $t_0$ on the initial local temperature $T^*$\ is observed in our calculation.
\section{Conclusion}\label{se_cl} 
Based on the rate equation of heavy quarkonia and the Bjorken's hydrodynamics, we calculated the gluon number density $f_{\omega}$ in unit energy and the loss term $\alpha_{J/\psi}$ in the finite space compared with that in the infinite space at the same temperature. It is found that the suppression of heavy quarkonia in the constrained space is weak, and therefore the suppression of heavy quarkonia at early time after the collision can be neglected, even if the temperature of the medium is high. The resulting concept effective initial time $t_0$ can be estimated at $p_T=0$ by the threshold energy, and it increases with $p_T$. A modified Euler-Maclaurin formula is developed to deal with the summation powerfully.
\section{Acknowledgments} 
The work is supported by the NSFC under the Grant No.s 11547043, 11705125 and by the ``Qinggu'' project of Tianjin University.
\section*{References} 
\bibliographystyle{elsarticle-num}
\bibliography{mybibfile}  
\end{document}